\documentclass[fleqn,usenatbib]{mnras}
\usepackage[utf8]{inputenc}
\usepackage{graphicx}
\usepackage{hyperref}
\usepackage{amssymb}
\usepackage{amsmath}
\usepackage{amsfonts}
\usepackage{multirow}
\usepackage{calc}
\usepackage{comment}
\usepackage{soul}
\usepackage{xcolor}
\usepackage{soul}

\newcommand{\mpcbyh}{\, h^{-1} \, \mathrm{Mpc}}

\newcommand{\MSUN}{{{M}}_{\odot}}
\newcommand{\lya}{\rm {Ly{\alpha}}}

\newcommand{\Omegam}{\Omega_{\rm m}}
\newcommand{\mFDM}{m_{22}}

\newcommand{\ffdm}{f_{\textup{F}}}

\newcommand{\kJeansEq}{k_{\textup{Jeq,FDM}}}
\newcommand{\MJeans}{M_{\textup{J,FDM}}}

\newcommand{\PCDM}{P_{\textup{CDM}}}
\newcommand{\PFDM}{P_{\textup{FDM}}}
\newcommand{\PDM}{P_{\textup{DM}}}
\newcommand{\TFDM}{T_{\textup{F}}}
\newcommand{\deltacrit}{\delta_{\textup{c}}}

\newcommand {\Scut}     {S_{\rm T}}

\title[FDM and high-$z$ HMF]{Fuzzy Dark Matter Halo Mass Functions at Cosmic Dawn}

\author[Ghara, Lidz, Grin, \& Sipple]{
Raghunath Ghara,$^{1,2}$\thanks{E-mail: ghara.raghunath@gmail.com}
Adam Lidz,$^{3}$
Daniel Grin$^{1}$
and Jackson Sipple$^{3}$
\\
$^{1}$Department  of  Physics  and  Astronomy,  Haverford  College, 370  Lancaster  Avenue,  Haverford,  Pennsylvania  19041,  USA\\
$^{2}$Department of Physics, Indian Institute of Technology Kharagpur, Kharagpur 721 302, India\\
$^{3}$Department of Physics and Astronomy, University of Pennsylvania, 209 South 33rd Street, Philadelphia, PA 19104, USA
}

\date{Accepted XXX. Received YYY; in original form ZZZ}

\pubyear{2026}

\begin{document}
\label{firstpage}
\pagerange{\pageref{firstpage}--\pageref{lastpage}}
\maketitle

% Abstract of the paper
\begin{abstract}
In fuzzy dark matter (FDM) cosmological models, wave effects impact astrophysical length scales, suppressing the abundance of small mass dark matter halos, and delaying the earliest phases of galaxy formation during Cosmic Dawn. Current and upcoming James Webb Space Telescope (JWST) measurements of the galaxy ultraviolet luminosity function (UVLF) will allow unprecedented tests of this suppression, yet significant uncertainties remain in theoretical models of the FDM halo mass function. We run a new suite of N-body simulations with FDM particle masses of $mc^{2}=10^{-22}\,{\rm eV} - 2 \times 10^{-21}$ eV and mixed FDM-cold dark matter (CDM) models with FDM mass fractions of $f_{\mathrm{F}} = 0.3-1$. We identify and remove spurious halos from discreteness noise and quantify the associated systematic uncertainty. We provide a new halo mass function fitting formula, calibrated over $z=6-11$, applicable to pure FDM and mixed dark matter scenarios.  Our results are in better agreement with previous simulation-based fitting formulas than with current semi-analytic mass function models. Nevertheless, for $m c^{2} = 10^{-21}$ eV and $M \sim 3 \times 10^9 M_\odot$ we find a $\sim 30\%$ weaker suppression than earlier simulation-based formulas predict, which we attribute to their extrapolation beyond the $m_{\rm FDM}$ range previously simulated. Applying our fitting formula to the UVLF, we find that upcoming JWST observations behind foreground lensing clusters, probing $M_{\rm UV} \gtrsim -13$ at $z \gtrsim 10$, will provide a powerful test of FDM and mixed dark matter models.  
\end{abstract}

%%%%%%%%%

%%%%%%%%%%%%%%%%%
\section{Introduction}
\label{sec:intro}
 
Most of the universe's matter does not couple to electromagnetism and is observed only through its gravitational influence. This component, {\it dark matter} (DM), plays a central role in cosmic structure formation, yet its physical nature is unknown. One intriguing possibility is fuzzy dark matter (FDM) \citep{2000PhRvL..85.1158H}, also known as wave DM, axion-like  particle (ALP) DM, ultra-light axion (ULA) DM, or Bose–Einstein condensate DM (see \citealt{2016PhR...643....1M, 2021A&ARv..29....7F, 2021ARA&A..59..247H} and references therein). In contrast to cold dark matter (CDM), FDM is composed of ultralight bosons with masses $m \sim 10^{-22}~{\rm eV}/c^2$. The corresponding de Broglie wavelengths span astrophysical length scales, suppressing the abundance of low mass DM halos and producing other small-scale signatures. On large scales, FDM behaves like CDM and successfully reproduces observations of the large-scale structure of the universe, while predicting testable small-scale departures from CDM (see e.g. \citealt{2016PhR...643....1M, 2020PrPNP.11303787N, 2021ARA&A..59..247H, Ohare24}).

FDM is well motivated by theories beyond the standard model of particle physics, such as string theory, which predicts ultra-light scalar and pseudo-scalar fields (the dilaton and axions, respectively, e.g. \citealt{Hui:2016ltb}). In \emph{axiverse} scenarios, many such fields exist, with masses spanning many orders of magnitude \citep{2010PhRvD..81l3530A, 2012JHEP...10..146C, 2026arXiv260223424B}. These fields could contribute significantly to the cosmic energy budget, yielding a FDM cosmology or a mixture of FDM and CDM.

Astrophysical observations at small scales are useful test of FDM and mixed FDM-CDM models. FDM constraints are also imposed at larger scales by CMB anisotropies \citep{2015PhRvD..91j3512H,2018MNRAS.476.3063H,2026arXiv260606410L}, galaxy clustering \citep{2023JCAP...06..023R}, and weak lensing \citep{2022MNRAS.515.5646D}. Small-scale FDM probes include the $\lya$ forest \citep{2017PhRvL.119c1302I, 2017PhRvD..96l3514K, 2017MNRAS.471.4606A, 2021PhRvL.126g1302R}, the abundance/dynamics of Milky Way (MW) satellites \citep{2020PhRvD.101l3026S, 2021PhRvL.126i1101N, 2021JCAP...10..043B,2025ApJ...986..127N}, the global 21-cm signal  \citep{2018PhRvD..98b3011L, 2019JCAP...04..051N}, galaxy ultraviolet luminosity functions (UVLFs) \citep{2016ApJ...818...89S, 2017PhRvD..95h3512C, 2019MNRAS.488.5551N, 2024ApJ...976...40W, 2025MNRAS.538.1830S}, disk heating \citep{2019ApJ...871...28B,2024MNRAS.530..129Y}, and DM streams \citep{2021JCAP...03..076D} in MW-type galaxies. 

In the case of the UVLF, the James Webb Space Telescope (JWST) is enabling unprecedented observations at redshifts of $z \gtrsim 9$ \citep{2023ApJS..265....5H, 2023MNRAS.523.1009B, 2023ApJS..265....5H,2024MNRAS.533.3222D,Atek26}, providing new opportunities to test FDM. Measurements at the highest accessible $z$ discriminate between FDM and CDM because CDM structure formation at early times is dominated by low-mass halos, which are mostly absent in FDM. Thus, FDM is expected to show a distinctive delay in the onset of structure formation, which can be tested at e.g. $z \gtrsim 9$ for halo masses of $M \lesssim 10^{10} M_\odot$. For pure FDM scenarios, a limit of $mc^{2}\gtrsim 1.5\times 10^{-21}~{\rm eV}$ was determined from UVLF measurements in the Hubble Frontier Fields \citep{2025MNRAS.538.1830S}.

Although the UVLF imposes a powerful constraint to FDM, significant modeling uncertainties remain. A key challenge is predicting the halo mass function (HMF), as is required to connect the UVLF with underlying DM halo populations. In principle, predicting FDM HMFs requires simulations of non-linear wave dynamics over cosmological volumes. This is computationally prohibitive for the viable FDM parameter space, as the FDM de Broglie wavelengths must be resolved in a large cosmological simulation box \citep[e.g.,][]{2014NatPh..10..496S, 2014PhRvL.113z1302S, 2019PhRvD..99f3509L, 2021MNRAS.506.2603M}. Many FDM simulations thus include small-scale power suppression in the initial conditions but neglect subsequent wave dynamical (e.g. \citealt{2016ApJ...818...89S}, henceforth S2016). Furthermore, determining the FDM HMF using simulations requires the removal of ``spurious halos'' resulting from discreteness noise in the initial conditions (S2016).\footnote{Note that the noise from the finite particle sampling in an N-body simulation departs from a Poisson distribution \citep{2007MNRAS.380...93W}. Thus we use the term ``discreteness noise.''}

Accurate and fast models of the HMF in pure and mixed FDM-CDM cosmologies are essential to interpret observations, requiring robust simulations and fitting formulae. The HMF has been simulated for mixed scenarios using full wave dynamics, but only for $mc^{2}=10^{-24.5}~{\rm eV} $ and FDM mass fractions of $\ffdm\leq 0.3$ \citep{2025MNRAS.537..252D,2026arXiv260606599J}. Fitting formulae were obtained, but not tested beyond this parameter range. FDM $N$-body simulations spanning a broader parameter space exist, but only for zoom-in simulations of the sub-halo mass function \citep{2025ApJ...986..127N}.

We thus run a suite of FDM $N$-body simulations with $\mFDM\equiv mc^{2}/10^{-22}~{\rm eV}=1,10,20$, and for mixed scenarios with $\mFDM=1$ and FDM fractions of $\ffdm=0.3, 0.5$, and $0.9$. These include suppressed small-scale initial conditions but not subsequent wave-dynamical effects. We cross check the FDM HMF fitting formula from S2016, pushing to $\mFDM$ values that are a factor of several larger than considered there. We quantify the systematic uncertainties in removing spurious halos. We present a HMF fitting formula for pure FDM and mixed FDM-CDM cosmologies and compare it to semi-analytic models. We then use our new fitting formula to model the UVLF, and discuss prospects for upcoming JWST measurements.

This paper is organised as follows. In Section \ref{sec:hmf}, we discuss the uncertainties associated with HMF predictions in FDM models. In Section \ref{sec:method}, we describe our simulations and methodology for estimating the HMF. We present results in Section \ref{sec:results}, followed by discussion in Section \ref{sec:dis} and conclusions in Section \ref{sec:con}. Throughout this paper, we use the cosmological parameters $\Omega_m = 0.309$, $\Omega_\Lambda = 0.691$, $h = 0.67$, $n_s=0.965$ and $\sigma_8 = 0.83$, consistent with \citet{Planck20}.

%%%%%%%%%%%%%%%%%%
\section{FDM Halo Mass Function Uncertainties}
\label{sec:hmf}

Before presenting our new simulation results, it is useful to contextualize these efforts by comparing HMF models in the current literature. We aim to better understand the relatively large range of FDM HMF predictions summarised below.

In addition to simulation-based fitting formulas, there are semi-analytic FDM HMF models based on the excursion set formalism. Although semi-analytic models of the CDM HMF provide intuition and broadly reproduce the results of simulations, scenarios with suppressed initial small-scale power, such as FDM \citep[e.g.,][]{2014MNRAS.437.2652M, 2017MNRAS.465..941D, 2022MNRAS.510.1425K} and warm DM (WDM), are more challenging to handle \citep{2013MNRAS.428.1774B, 2013MNRAS.433.1573S,2018JCAP...04..010L}. In particular, HMF predictions in these scenarios are sensitive to the choice of smoothing filter.

The often employed real-space top-hat filter does not lead to a strong suppression in the low-mass HMF, in stark contrast to simulations. Sharp $k$-space filters are thus used (e.g. \citealt{2022MNRAS.510.1425K}) to obtain suppressed HMFs, but these lead to an ambiguous mapping from filtering scale to halo mass \citep{2013MNRAS.428.1774B}, as discussed in Appendix A. Other work has shown that in models with suppressed-small scale power (which can be caused by non-standard inflationary models or a variety of DM candidates), a wide range of smoothing filters can lead to vastly different predictions for the HMF \citep{2013MNRAS.433.1573S,2018JCAP...04..010L}.

The usual excursion set formalism for the HMF is well suited to CDM, where structure formation is hierarchical, but may break down in FDM, where the rms density fluctuations become nearly mass independent at low halo masses. The dynamical effects of FDM may also play a role in determining the FDM HMF, which has been approximated through the use of halo mass-dependent linear overdensity collapse thresholds \citep{2014MNRAS.437.2652M,2016arXiv160505973M}. The resulting HMF also depends on whether the computation of random walk statistics include the scale-dependence of barrier crossing \citep{2017MNRAS.465..941D}, with a variety of relevant techniques also discussed in \citet{2006ApJ...641..641Z,2009ApJ...696..636R,2013MNRAS.433.3428F}.

Wave-dynamical FDM simulations are currently only feasible for $\mFDM\equiv mc^{2}/10^{-22}~{\rm eV} \leq 0.7$, and in small ($\sim 10 \mpcbyh$) simulation boxes \citep{2023MNRAS.524.4256M}. Those results for the HMF are broadly consistent with earlier (e.g. S2016) $N$-body simulation work, which only included the cutoff in the initial power spectrum but not subsequent wave dynamics.  This motivates our simpler and computationally less expensive $N$-body approach for determining the HMF.

\begin{figure}
 \centering
\includegraphics[width=\columnwidth]{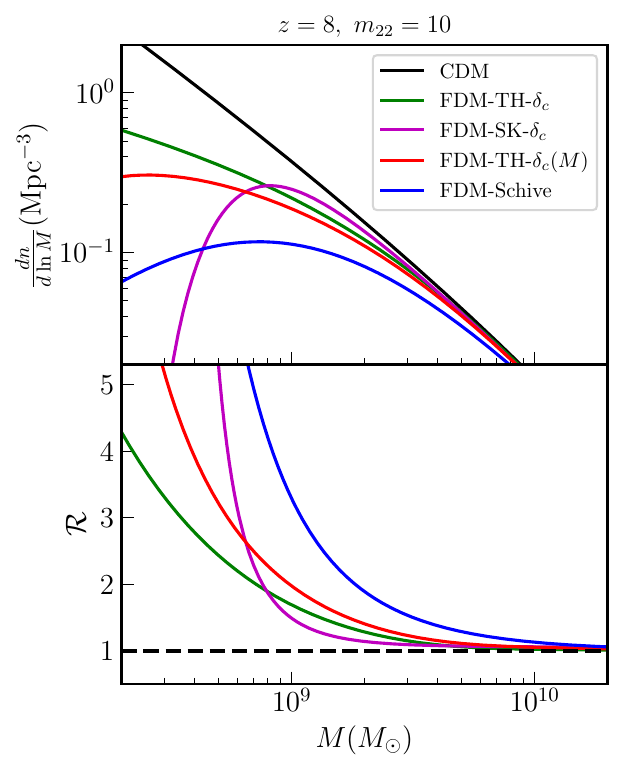}
    \caption{Comparison of different HMFs in an FDM model with $\mFDM=10$ at $z=8$. The top panel shows the FDM HMFs computed using different prescriptions, along with the corresponding CDM HMF. The bottom panel displays the ratio of the CDM to FDM HMFs as a function of halo mass. See Appendix \ref{sec:allhmf} for detailed descriptions of each model.}
   \label{image_hmfs}
\end{figure}

We compare the HMF model variants discussed above for an example case with $\mFDM=10$ and $z=8$ in Figure \ref{image_hmfs}. We adopt the CDM HMF formula from \citet{2002MNRAS.329...61S}, which is motivated by ellipsoidal collapse extensions to excursion set theory with a mass-dependent collapse barrier. 

The specific models considered here are: (1) the FDM HMF calculated using a real-space top-hat filter (FDM-TH-$\delta_c$) \citep[e.g.,][]{1974ApJ...187..425P,2002MNRAS.329...61S}. In this case, the functional form is assumed to follow the \citet{2002MNRAS.329...61S} results, with FDM effects entering only through the modified halo mass dependence of the linear variance,
(2) the FDM HMF obtained using a real-space top-hat filter with a mass-dependent barrier intended to approximately account for  modified linear growth in FDM [FDM-TH-$\delta_c(M)$] \citep[e.g.][]{2014MNRAS.437.2652M}, (3) the FDM HMF determined using a sharp $k$-space window function (FDM-SK-$\delta_{c}$) \citep[e.g.][]{2022MNRAS.510.1425K}, and (4) the HMF fitting formula to the $N$-body simulations of \citet[][]{2016ApJ...818...89S}. Appendix \ref{sec:allhmf} describes each model in detail.

Although each FDM HMF prescription yields a suppression (compared to CDM) for low halo masses, the models differ significantly in the level and shape of the suppression. The S2016 fitting formula gives the strongest suppression across most of the halo mass range shown. Since this formula is based on simulations which include only the cut-off in the initial power spectrum, one might expect the FDM-TH-$\delta_c(M)$ case -- which approximately accounts for FDM dynamical effects -- to show a lower halo abundance. Instead, the higher abundance in the semi-analytic model may indicate issues with the model and/or the simulation-based fitting formula. Note that over the mass range shown, the FDM-TH-$\delta_c$ and FDM-TH-$\delta_c(M)$ models show a similar shape with a gentle decline. However, the FDM-TH-$\delta_c(M)$ HMF has a steep cut-off at smaller masses $M\lesssim 4\times 10^8\, \MSUN$. Finally, the sharp-$k$ HMF suppression sets in at lower masses but has a more rapid decline than in the S2016 model. 

The primary takeaway from Fig.~\ref{image_hmfs} is that there are substantial theoretical uncertainties in these predictions. The spread among the semi-analytic HMF models reflects differences in their filtering schemes and treatments of the collapse barrier. In the case of simulations, there are uncertainties related to the treatment of FDM dynamics, the removal of spurious halos, and the extrapolation of the S2016 results beyond the $\mFDM$ range over which they were calibrated. As an example of the current spread across models, the ratio of CDM to FDM HMFs ($\mathcal{R}$) for $z=8$ and $\mFDM=10$ at $10^9 \, \MSUN$ lies in the range $\approx 1.4-3$. The minimum value of $\mathcal{R}$ is obtained for the FDM-SK HMF, while the FDM-Schive HMF yields the maximum. These uncertainties motivate a reexamination of the halo abundance predictions using the new suite of FDM N-body simulations presented here. 
   
%%%%%%%%%%%%%%%%%%%%%%
\section{Methodology}
\label{sec:method}

In this section, we describe the $N$-body simulations run for this study and our approach for identifying and excising spurious halos. Our methodology generally follows that of S2016, with some differences discussed below. 

%%%%%%%%%%%%%%%%
\subsection{$N$-body simulations}
\label{sec:nbody}

We carry out dark-matter only $N$-body simulations using the publicly available {\sc gadget2} code \citep{2005MNRAS.364.1105S}. We use a modified version of the {\sc N-genIC} code to generate initial conditions at $z=100$ \citep{2005Natur.435..629S} in FDM and mixed scenarios using suitable transfer functions. Denoting the mass fraction of FDM by $\ffdm$ (with a fraction $1 - \ffdm$ in CDM), the linear matter power spectrum at redshift $z$ is
\begin{equation}
\begin{aligned}
\PDM(k,z) &= \left[\ffdm^2\,\TFDM^2(k)
+2\ffdm(1-\ffdm)\,\TFDM(k)
+(1-\ffdm)^2\right] \\
&\times \PCDM(k,z), \\
\TFDM(k) &\simeq \frac{\cos(x^3)}{1+x^8}, \\
x &\equiv 1.61\,\mFDM^{1/18}\left(\frac{k}{\kJeansEq}\right),
\label{EQ_MIXFDM_PS}
\end{aligned}
\end{equation}
where $\PDM$ is the total power spectrum of the FDM-CDM mixture, while $\PCDM$ is the CDM power spectrum according to the transfer function of \citet{1998ApJ...496..605E}. Here $\TFDM$ is the FDM transfer function from \cite{2000PhRvL..85.1158H}, while $\kJeansEq = 9\,\mFDM^{1/2}\,{\rm Mpc}^{-1}$ is the FDM Jeans wavenumber at matter–radiation equality. The quantity $\mFDM$ is the mass of the FDM particles in units of $10^{-22}~{\rm eV}/c^2$. This transfer function is used for consistency and comparison with S2016.\footnote{We note that more accurate transfer functions could be obtained using the \textsc{AxieCAMB} code developed in \citet{2025PhRvD.112b3513L}, \href{https://github.com/Ra-yne/AxiECAMB}{https://github.com/Ra-yne/AxiECAMB}.} The FDM transfer function $\TFDM$ describes the small-scale suppression in the FDM fluctuations (relative to CDM), while the middle term in the square brackets of Equation~\ref{EQ_MIXFDM_PS} accounts for the cross-power between the CDM and FDM fluctuations in linear theory. The $\ffdm=0$ limit is a pure CDM model, while $\ffdm=1$ is a pure FDM scenario.

\begin{figure}
 \centering
\includegraphics[width=\columnwidth]{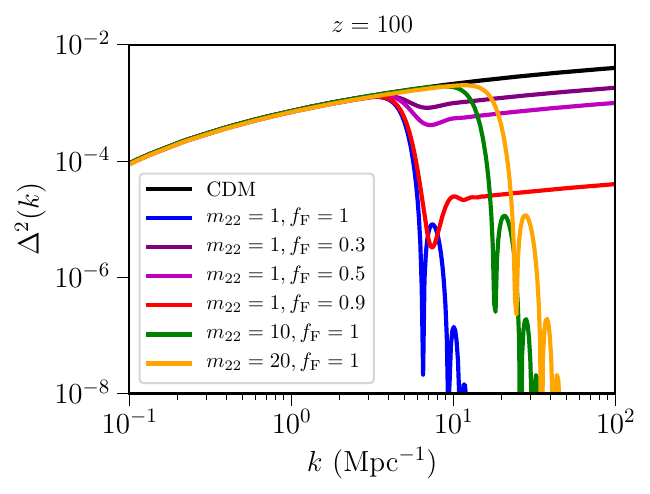}
    \caption{Dimensionless power spectra $\Delta^2(k)=k^3P(k)/2\pi^2$ at $z=100$ for the different dark matter scenarios considered in this study.  The suppression in the initial power spectrum from FDM moves to smaller scales (higher $k$) with increasing $\mFDM$, while the fluctuations are less strongly suppressed as the FDM fraction $\ffdm$ decreases. The initial conditions for our $N$-body simulations follow these power spectrum models.}
   \label{image_psk}
\end{figure}

Fig.~\ref{image_psk} shows the linear matter power spectra at $z=100$ for the various DM scenarios explored in this study. As expected, the FDM spectrum exhibits a sharp cutoff near $k \sim \kJeansEq$, followed by oscillatory features at higher $k$. In contrast, in the mixed cases the CDM component still clusters strongly on small scales, and hence the mixed scenarios show less suppression at $k \gtrsim \kJeansEq$ than in pure FDM models. 

For each of the DM models explored here, we run simulations with a comoving box length of $20 \mpcbyh$ and particle numbers of $512^3$ and $1024^3$. Although this relatively small box~size is insufficient for capturing the abundance of rare massive halos, our main objective in this work is to determine the ratio of the CDM to FDM HMFs at low masses. While finite box size effects can also impact estimates of the HMF in this regime, the {\em ratio} of the CDM to FDM HMF should still be fairly robust, since sample variance effects are expected to largely cancel when taking ratios. 

The $N$-body particle mass is $2 \times 10^6 \, \MSUN$ in our high-resolution runs, while it is $1.7\times 10^7\,\MSUN$ in the lower resolution simulations. These simulations have sufficient resolution to study halos near the suppression mass of roughly $M \sim 10^9 M_\odot$ for $\mFDM \sim 10$. Comparing simulations at these two resolutions is important for identifying spurious halos and testing our results.  

\begin{table}
\centering
\caption{Summary of N-body simulations. All runs use a comoving box length of $20\,h^{-1}$\,Mpc. The N-body particle mass is $m_p = 2\times10^6\,M_\odot$ for high-resolution runs ($1024^3$ particles) and $m_p = 1.7\times10^7\,M_\odot$ for low-resolution runs ($512^3$ particles). Snapshots are stored across
$z = 6$--$15$ for all runs.}
\label{tab:sims}
\begin{tabular}{l c c c}
\hline\hline
\textbf{Simulations} & $\mFDM$ & $f_{\rm FDM}$ & $N_{\rm part}$ \\
\hline
CDM reference & --- & 0.0 & $512^3,\,1024^3$ \\
\hline
\multirow{3}{*}{Pure FDM \hspace{1.2cm}\Bigg\{}
& 1  & 1.0 & $512^3,\,1024^3$ \\
& 10 & 1.0 & $512^3,\,1024^3$ \\
& 20 & 1.0 & $512^3,\,1024^3$ \\
\hline
\multirow{3}{*}{Mixed FDM--CDM\hspace{0.22cm}\Bigg\{ }
& 1 & 0.3 & $512^3,\,1024^3$ \\
& 1 & 0.5 & $512^3,\,1024^3$ \\
& 1 & 0.9 & $512^3,\,1024^3$ \\
\hline\hline
\end{tabular}
\end{table}

In total, we run 14 different simulations, spanning variations in FDM parameters and resolutions. First, we run a pure CDM reference model. Next, we consider pure ($\ffdm=1$) FDM scenarios with $\mFDM = 1, 10, 20$. Since each model is run at two resolutions, the pure CDM and pure FDM models amount to 8 simulation runs. We further investigate mixed scenarios with $\mFDM = 1$ and each of $\ffdm = 0.3, 0.5, 0.9$ (6 runs). The particle positions are stored for a series of simulation snapshots, spanning the redshift range between $z=6$ and $z=15$. The HMFs are estimated in a subsequent post-processing step from stored snapshots. The simulation runs are summarized in Table~\ref{tab:sims}.

%%%%%%%%%%%%%%%%%%%%%%%%%%%%%%%%%%%%%%%%%%%%%%%%%
\subsection{Estimating Halo Mass Functions}

We use the AMIGA Halo Finder \citep[AHF,][]{2009ApJS..182..608K} to identify DM halos from the stored particle positions in our {\sc gadget2} simulations. The AHF uses an adaptive mesh refinement grid to locate density peaks and a spherical overdensity method to identify bound particles. The AHF code also returns particle IDs for the DM particles within each simulated halo. The particle IDs help to identify and remove spurious halos. 
 
As mentioned in the introduction, spurious halo contamination is a concern for FDM simulations. Spurious halos arise from particle discreteness noise, which becomes the dominant source of small-scale fluctuations in the initial conditions due to the high-$k$ cut-off in the FDM linear density power spectrum \citep[see e.g.,][]{2007MNRAS.380...93W, 2016MNRAS.462..474P}. As large-scale filaments form, this discreteness noise seeds artificial overdensities along the filaments that eventually collapse into spurious halos. 

Fortunately, the proto-halo regions of spurious halos have distinctive shapes. They typically originate from highly elongated structures in the initial conditions with extreme axis ratios. 
This contrasts with genuine halos, which originate from peaks in the primordial density fluctuation field and trace back to compact, well-connected Lagrangian regions with relatively moderate axis ratios. Moreover, genuine halos show stability with respect to changes in simulation resolution, whereas spurious halos do not. In particular, the overlap of proto-halo volumes identified at different resolutions is significantly smaller for spurious halos than for genuine ones.

We use these properties to identify spurious halos in the AHF catalog, following the approach of S2016 and earlier WDM simulation work \citep{2014MNRAS.439..300L}. First, we trace each halo particle back to its proto-halo region in the initial conditions at $z=100$. We then apply two criteria to determine whether a halo is robust or spurious.

The first criterion is based on the sphericity of the proto-halo regions, since spurious halos tend to originate from highly elongated Lagrangian structures. We define the sphericity coefficient $S$ of a proto-halo following S2016 as:
\begin{equation}
S = \sqrt{\frac{I_1+I_2-I_3}{-I_1+I_2+I_3}},
\label{eq_Sphericity}
\end{equation}
where $I_1, I_2$ and $I_3$ are the principal moments of inertia of a proto-halo, ordered such that $I_1 \le I_2 \le I_3$. In this way, we determine the initial (Lagrangian) sphericities of all simulated DM halos. Next, one must choose a threshold sphericity, $\Scut$, where halos with proto-halo sphericity $S \geq S_\mathrm{T}$ are considered robust while those with $S < \Scut$ are considered spurious. There is some ambiguity in the appropriate choice of $\Scut$, as discussed below. 

\begin{figure}
 \centering
\includegraphics[width=\columnwidth]{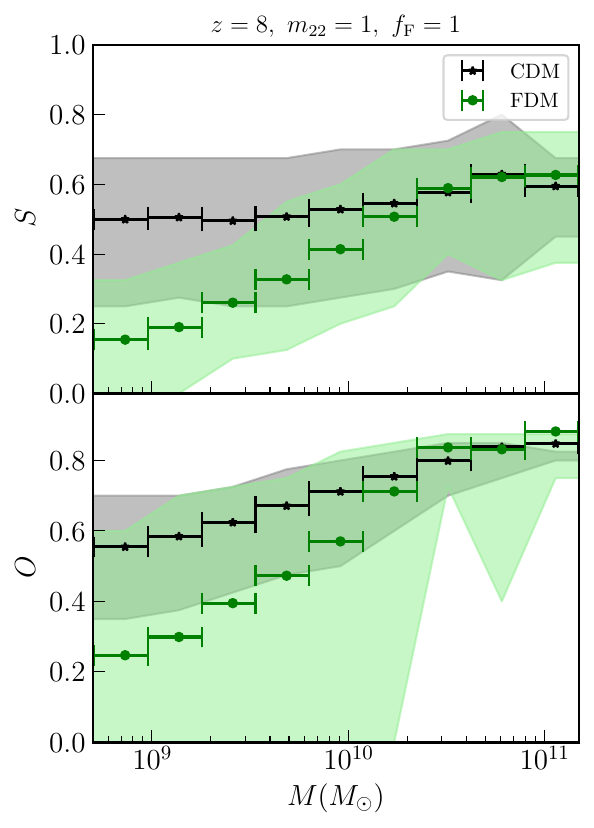}
    \caption{The proto-halo sphericity (top) and overlap (bottom) coefficients as a function of halo mass at $z=8$. The black points and error bars show the CDM results, while the green ones show the same for an FDM model with $\mFDM=1$ and $\ffdm=1$. The shaded regions show the $2\sigma$ spread. Note that the simulated halo abundance is small at $M \gtrsim 10^{10.5} M_\odot$, leading to noise in the shaded boundaries. Low mass FDM halos often show smaller $S$ and $O$ coefficients than in CDM, since halos with small sphericity and/or overlap coefficients are frequently artefacts of simulation discreteness noise. We remove halos with small $S$ or $O$ as spurious.}
   \label{image_S_post}
\end{figure} 

The second criterion compares low and high-resolution simulations for the same FDM model and quantifies the matching accuracy between corresponding proto-halos in those simulations. The spatial overlap coefficient $O$ can be defined as
\begin{equation}
O = \frac{ V_{\rm low} \cap V_{\rm high} } { V_{\rm low} \cup V_{\rm high} },
\label{eq:Overlap}
\end{equation}
where $V_{\rm low}$ and $V_{\rm high}$ are the volumes of the physical space occupied by the same proto-halo in the low and high-resolution simulations, respectively. Thus, $O$ lies between $[0,1]$ with $O=1$ indicating that the volumes spanned by a proto-halo region overlap perfectly, while $O=0$ indicates no overlap. Here, low and high-resolution simulations refer to the $512^3$ and $1024^3$ particle simulations, respectively. We use the cloud-in-cell scheme for estimating these volumes by depositing proto-halo particles onto regular grids, with the separation between grid cells matching the mean inter-particle spacing. Physical halos should not appear/disappear when the simulation resolution changes, and should thus have high $O$ values.

In order to determine fiducial threshold choices, $\Scut$ and $O_\mathrm{T}$, we first consider the distribution of the $S$ and $O$ coefficients in our CDM halo catalogs. Here and throughout we consider halos with at least 30 DM $N$-body particles, corresponding to a minimum halo mass of $5 \times 10^8 \MSUN$ in our $512^3$ particle simulations. The CDM $S$ and $O$ proto-halo coefficients for halos at $z=8$ are shown in Fig.~\ref{image_S_post}, where the shaded regions span the 2-$\sigma$ (95\%) range in the simulated halos. Since we expect the CDM halos to be robust, we choose $\Scut$ and $O_\mathrm{T}$ such that 95\% of the simulated CDM halos in the lowest halo mass bin lie above each threshold. 
At $z=8$, these criteria yield $\Scut=0.25$ and $O_\mathrm{T}=0.3$. 

We find that the threshold values, according to the 95\% criterion, remain similar across $z=6-11$, the main redshift range considered in this work. We therefore adopt $\Scut=0.25$ and $O_\mathrm{T}=0.3$ as fiducial values across all redshifts, but explore the sensitivity of our results to these choices in what follows (in particular, in Fig.~\ref{image_SO-thre-full}). We note that our estimates of $S_\mathrm{T}$ and $O_\mathrm{T}$ are similar to, but slightly different from, those of S2016,  $S_\mathrm{T, Sch}=0.31$, and $O_\mathrm{T, Sch}=0.34$.  
 
We show the sphericity and overlap coefficients for CDM (black points) and an FDM model with $\mFDM=1$ and $\ffdm=1$ (green points) at $z=8$ in Fig.~\ref{image_S_post}. As anticipated, the FDM halos show smaller $S$ and $O$ on average than in CDM, and a wider spread in coefficients. This is especially the case at low halo mass. This suggests that a significant fraction of the $M \sim 10^9 M_\odot$ halos in our $\mFDM=1$ model are artefacts of discreteness noise. We will therefore remove them from our halo catalogs based on the $\Scut$ and $O_\mathrm{T}$ thresholds before estimating the HMFs. 

%%%%%%%%%%%%%%%%%%%%%%%%%
\section{Results}
\label{sec:results}

\begin{figure}
 \centering
\includegraphics[width=\columnwidth]{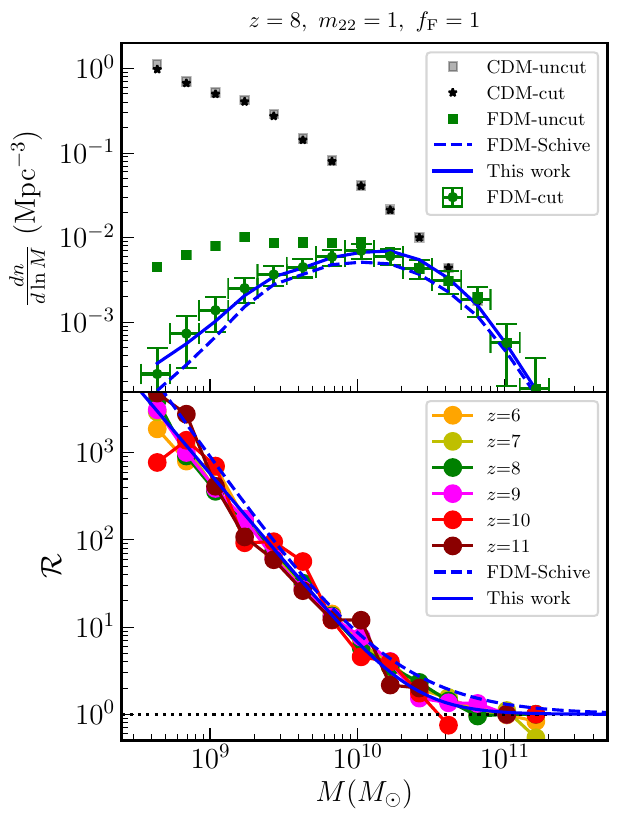}
    \caption{The top panel shows the $z=8$ CDM and FDM HMFs for $\mFDM=1$. The black stars show the CDM HMF after removing dark matter halos with $S$ and $O$ coefficients smaller than $S_\mathrm{T}$ and $O_\mathrm{T}$, respectively. The grey squares are the (nearly identical) CDM results without excising halos. The green squares give the FDM HMF without removing halos, while the green circles show the FDM HMF (mean and 1$\sigma$ Poisson error bars) after removing spurious halos using our fiducial $\Scut$ and $O_\mathrm{T}$ cuts. The solid (dashed) blue curves show the HMF according to our (Schive's) fitting formula (S2016). The bottom panel gives the ratio between the CDM and FDM HMF at $z=6$ to $z=11$ after removing spurious halos, compared to the fitting formulae. The simulated ratios are redshift-independent to within uncertainties, and are similar to those of S2016.}
   \label{image_fid}
\end{figure} 

\begin{figure}
 \centering
\includegraphics[width=\columnwidth]{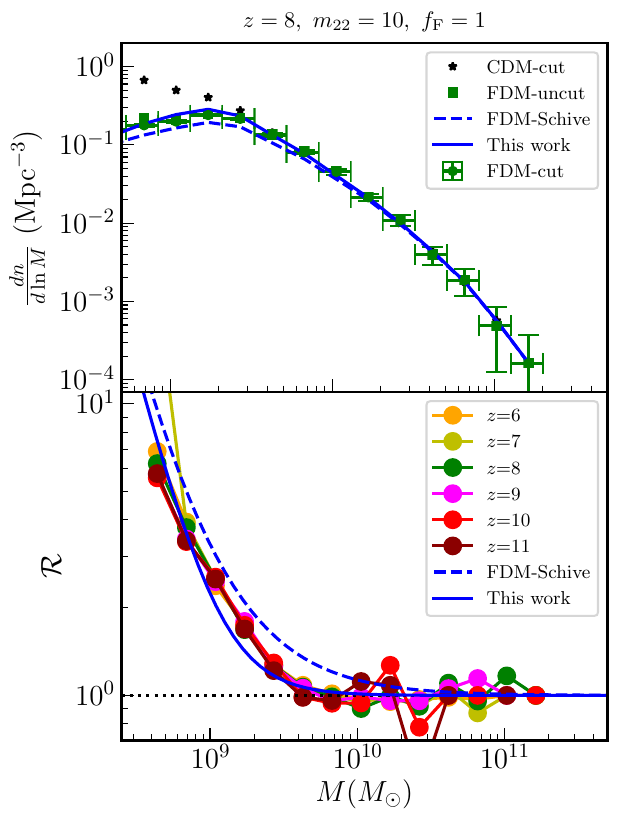}
    \caption{Same as Fig.~\ref{image_fid} but for $\mFDM=10$ and $\ffdm=1$. Again, the results are consistent with a redshift independent ratio. Here our results show a slightly weaker suppression than expected from the S2016 formula. Note that the simulation results are somewhat noisy above $M \gtrsim 10^{10.5} M_\odot$, particularly at $z \sim 11$.}
   \label{image_m10}
\end{figure}

\begin{figure}
 \centering
\includegraphics[width=\columnwidth]{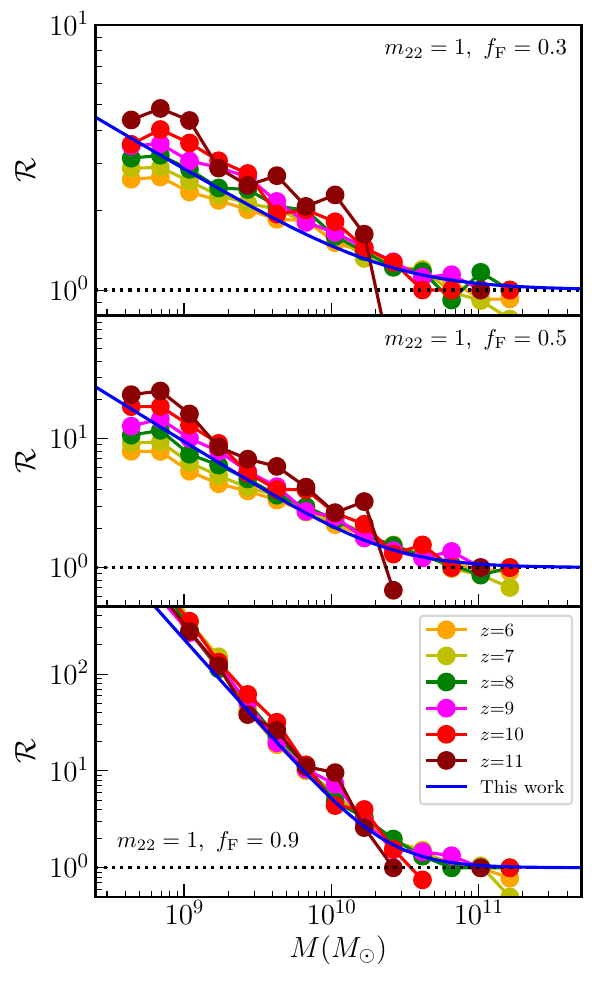}
     \caption{Same as Fig.~\ref{image_fid} but for mixed FDM-CDM scenarios with $\mFDM=1$ and $\ffdm=0.3, 0.5$ and 0.9. The $z=11$ results are a bit noisy since halos are rare objects at this redshift.}
   \label{image_m1f-new}
\end{figure} 

Here, we present the FDM HMFs for the different FDM scenarios considered in this paper. We quantify the ratio between the CDM and FDM HMFs, and find a compact fitting formula describing the simulation results. We extend earlier work (S2016), to find a formula that applies to both pure
and mixed DM models. 

%%%%%%%%%%%
\subsection{Pure FDM Scenarios}

We first calculate the HMF in CDM and in pure FDM scenarios (i.e., $\ffdm=1$) with each of $\mFDM = 1, 10$, and 20. Here, one key objective is to check the earlier S2016 analysis, which covered $\mFDM = 0.8-3.2$, in the current observationally viable regime. As current empirical bounds generally suggest $\mFDM \gtrsim 10$, our higher $\mFDM = 10, 20$ simulations can test whether the S2016 fitting formula applies beyond the $\mFDM$ range over which it was directly calibrated. 

We first consider an FDM model with $\mFDM=1$. Although this scenario is strongly disfavored by current data, it is a useful test case given the strong HMF suppression in this model. It also lies within the $\mFDM$ range simulated previously (S2016), allowing more direct cross-checks with earlier work. 

We compare the simulated $z=8$ FDM HMF to that from our CDM simulation in Fig.~\ref{image_fid}. The green squares in the top panel show the FDM HMF before spurious halo removal, while the green circles show estimates after removing halos based on their $S$ and $O$ coefficients. The same cuts are applied to the CDM HMF (black stars), although the impact is small for CDM, as can be discerned by comparing the grey squares with the black stars. The FDM HMF shows the expected strong suppression at low halo mass, with visible departures from CDM at $M\lesssim 5\times 10^{10}\, \MSUN$. Before excising the spurious halos, the FDM HMF shows a fairly flat trend in halo mass below $M\lesssim 5\times 10^{10}\, \MSUN$. After removing the suspect halos, however, the HMF continues to drop towards low masses. 

We find that the following fitting formula successfully describes the results of our simulation suite, across all pure FDM cases as well as the FDM-CDM mixed scenarios:
\begin{equation}
 \left.  \frac{dn(M,z, \mFDM, \ffdm)}{d\ln M}\right|_{\rm FDM} \simeq\left. \frac{dn(M,z)}{d\ln M}\right|_{\rm CDM} \left[1+\left(\frac{M}{M_0}\right)^{\alpha}\right]^{\beta},
     \label{eq:Rfitting}
 \end{equation}
 with $\alpha=-2\sqrt{\ffdm}$, $\beta=-\ffdm$, $M_0=M_\mathrm{cut} \, \mFDM^{-4/3} $ and $M_\mathrm{cut}=2.4\times10^{10} \, M_\odot$. Here, $\left.dn(M,z)/d\ln M\right|_{\rm CDM}$ corresponds to the CDM HMF estimated from our $N$-body simulations. The functional form of Equation \ref{eq:Rfitting} is similar to that in S2016, but adapted to treat mixed case scenarios and to better fit the results of our new simulation suite. The best-fit parameter values have been determined based on our simulation results at $z=6-11$, across all models. As discussed further below, our fitting formula in pure FDM cases gives slightly different results than that of S2016, which is only calibrated at relatively small $\mFDM \leq 3.2$. Note that the dependence on $\ffdm$ is determined empirically, while ensuring that the formula recovers CDM in the limit $\ffdm \rightarrow 0$ and our pure FDM results in the $\ffdm \rightarrow 1$ limit.

We show the ratio of the CDM and FDM HMFs, $\mathcal{R}$, at different redshifts for $\mFDM=1$ in the bottom panel of Fig.~\ref{image_fid}. The results are consistent with a redshift-independent ratio across the full redshift range explored, $z \sim 6-11$, to within the noise of the simulation measurements. This is consistent with the redshift independent functional form of Equation~\ref{eq:Rfitting} and the results of S2016. 

Next, we explore pure FDM models with $\mFDM>1$. We consider two scenarios, one with $\mFDM=10$ and the other with $\mFDM=20$. We show the results for the $\mFDM=10$ case in Fig.~\ref{image_m10}. In this scenario, the deviation between the FDM and CDM HMFs sets in at smaller mass scales, as expected, with deviations occurring for halos with masses of roughly $M \lesssim 3 \times 10^9 ~\MSUN$. In this case, the spurious halo contamination is less pronounced across the halo mass range resolved in our simulations, although spurious halos would still be an issue at higher resolution/smaller mass scales.  

We see in the bottom panel of Fig.~\ref{image_m10} that the results are redshift independent, as in the $\mFDM=1$ case. Our refined fitting formula provides a good overall match to the simulation results. We find a slightly less pronounced HMF suppression than expected from the fitting formula in S2016. Quantitatively, for a representative halo mass of $M=3 \times 10^9 M_\odot$ and $\mFDM=10$, we find 30\% more DM halos than in the S2016 formula.

Although our fitting formula visually tracks the simulation data points across all redshifts and masses more faithfully than the S2016 model, quantifying the goodness-of-fit in each case requires accounting for the full HMF covariance matrix across different mass and redshift bins. A simple Poisson error bar estimate is unreliable given the spatial correlations between halos and their relatively high abundance in the lower mass bins. Future work will therefore be required to estimate the full covariance matrix using many simulation realizations and/or a jackknife or bootstrapping technique.

The results are qualitatively similar in the $\mFDM=20$ model. In this case, the HMF suppression relative to CDM sets in around $M \lesssim 10^9 ~\MSUN$ (see Fig.~\ref{image_mall}). Here, our simulation results again show a slightly weaker suppression than in S2016: at a mass scale of $M = 3 \times 10^9 M_\odot$, $\mathcal{R} = 1$ in our simulations (Fig.~\ref{image_mall}), while the fitting formula from S2016 gives a small 14\% suppression. The small differences between our results with S2016 most likely result from our simulations spanning larger $\mFDM$ values, given that our results agree well for $\mFDM=1$.

%%%%%%%%%%%%%%%%%%%%
\subsection{Mixed FDM scenarios}

In Fig.~\ref{image_m1f-new}, we show the HMF ratios, $\mathcal{R}$, for our mixed FDM-CDM scenarios with FDM mass fractions of $\ffdm = 0.3, 0.5$ and $0.9$, each assuming $\mFDM =1$. As expected, the HMF suppression is reduced with decreasing $\ffdm$ and $\mathcal{R}$ also rises less steeply towards small halo mass. Interestingly, near the low-mass end of the halos captured in our current simulations, the mixed scenarios show a hint of non-zero redshift evolution. Specifically, the HMF appears slightly more suppressed at the highest redshifts shown than at the lowest redshifts in the figure. 

It is unclear whether the hint of redshift evolution in mixed scenarios has a physical origin or is a numerical artefact. Physically, the mixed case transfer function includes power on small scales that may help progressively seed small-mass halos below the FDM suppression mass scale as the universe evolves. This may naturally lead to a smaller $\mathcal{R}$ at lower redshifts. On the other hand, the hint for redshift evolution is partly driven by the $z \sim 11$ simulation data where our halo sample is relatively small. The low-mass end is also where spurious halo contamination and finite resolution may play a role. It may be interesting to explore this further in future work. 

Recently, \citet{2025MNRAS.537..252D} modeled mixed DM with wave-dynamical simulations in cases where $mc^{2}=10^{-24.5}~{\rm eV}$ (a factor of $\sim 300$ smaller than those considered here), working in a parameter space where $\ffdm \leq 0.3$ and focusing on $z \leq 4$. In that work, halo models are developed to model low $z$ precision large-scale measurements, while our work targets the Cosmic Dawn UVLF with substantially larger $\mFDM$.

If we extrapolate their fitting formula to our regime, with $\mFDM = 1$ and $\ffdm=0.3$, we find a much stronger suppression than predicted by Equation \ref{eq:Rfitting}. For example, their formula yields a factor of $5.6$ fewer halos at $M=10^{10} M_\odot$, with the difference growing to a factor of $\sim 20$ by $M = 3 \times 10^9 M_\odot$. These discrepancies are likely caused by using that fitting formula well outside the parameter space in which it was calibrated. In future work, it would be interesting to simulate intermediate $\mFDM$ mixed DM scenarios, in order to cover the transition between the two fitting formulas. Our results may be safely applied to predict the Cosmic Dawn UVLF for interesting and viable mixed scenarios with $\mFDM \sim \mathcal{O}(1-10)$ and $\ffdm \sim 0.1-1$.

\begin{figure}
 \centering
\includegraphics[width=\columnwidth]{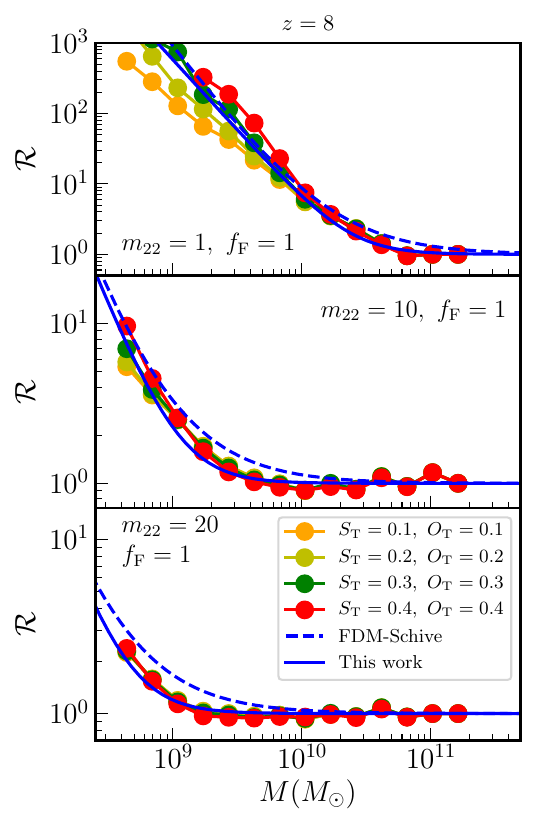}
    \caption{Ratio of the CDM and FDM HMFs ($\mathcal{R}$) as a function of halo mass for different combinations of $S_{\rm T}$ and $O_{\rm T}$ in pure FDM models with $\mFDM = 1$ (top), $\mFDM = 10$ (middle), and $\mFDM = 20$ (bottom) at a representative redshift of $z=8$. The solid blue lines show our fitting formula, while the dashed blue are from S2016.}
   \label{image_SO-thre-full}
\end{figure} 

\begin{figure}
 \centering
\includegraphics[width=\columnwidth]{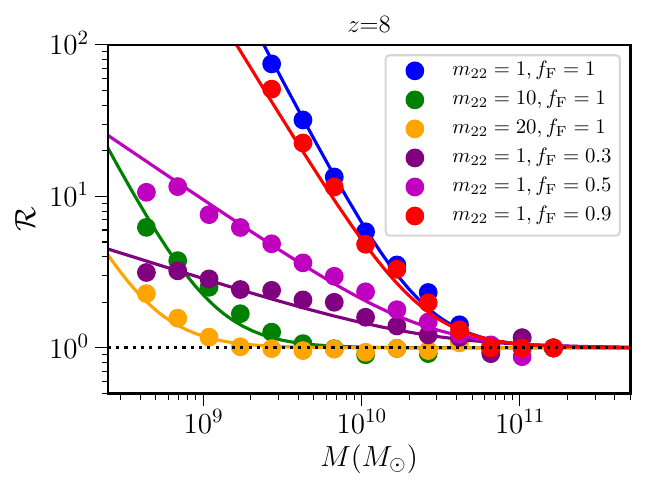}
    \caption{The ratio of CDM and FDM HMFs at $z=8$ for the full range of FDM scenarios considered in this work. The solid curves establish that our new fitting formula successfully describes this full range of models.}
   \label{image_mall}
\end{figure} 

\begin{figure*}
 \centering
\includegraphics[width=\textwidth]{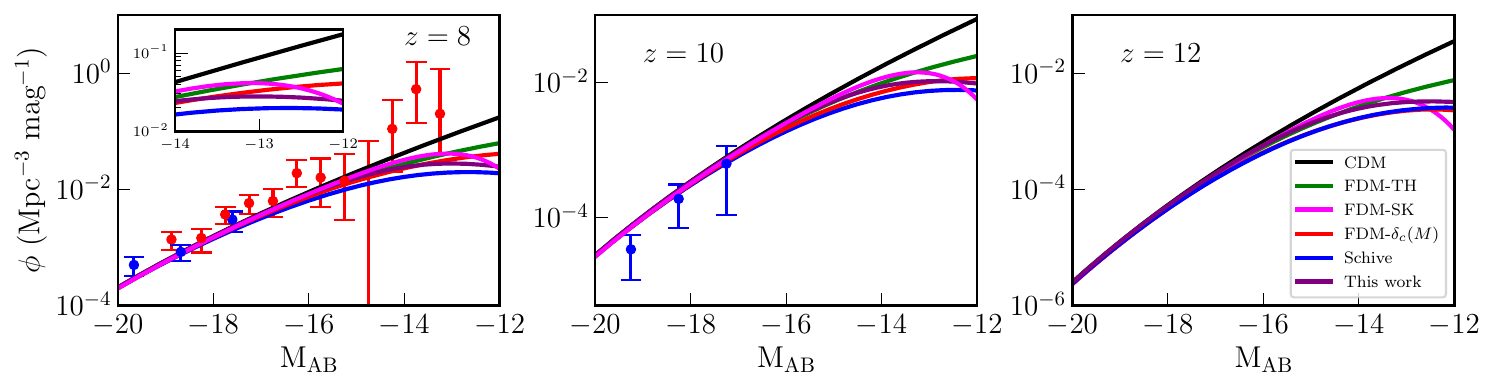}
    \caption{UVLF models and current data for $\mFDM=5$ and $\ffdm=1$ at $z= 8$, 10 and 12. The black curve assumes the CDM HMF, while other curves are for different FDM HMF/UVLF models. The points with $1\sigma$ error bars show UVLF measurements from \citet{2021AJ....162...47B} (blue, field galaxies) and \citet{2022ApJ...940...55B} (red, lensed galaxies). Our new HMF model should be used to refine previous FDM constraints from the UVLFs and to compare with upcoming JWST data.}
   \label{image_LF}
\end{figure*} 

\begin{figure*}
 \centering
\includegraphics[width=\textwidth]{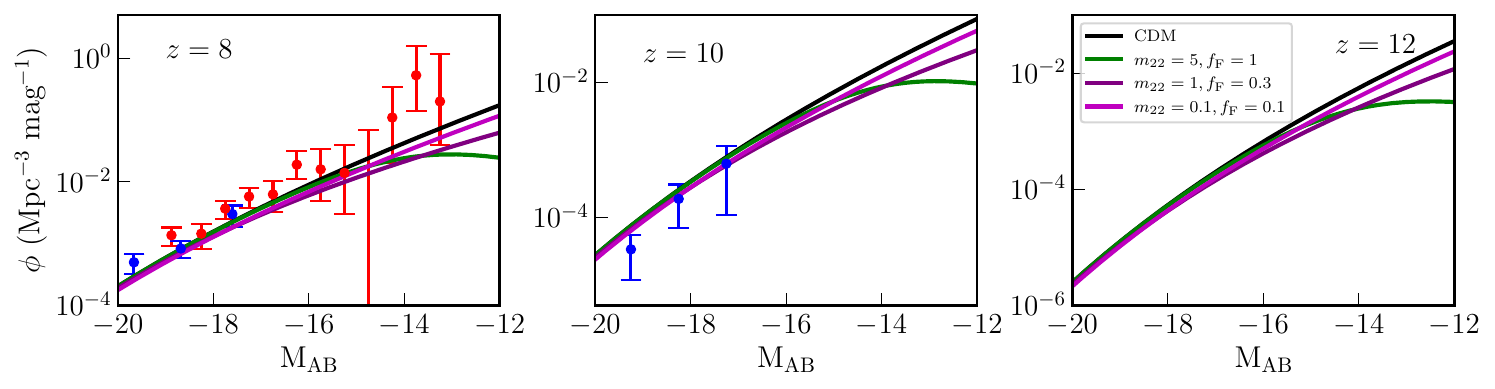}
    \caption{UVLF models and data at $z=8-12$ for mixed FDM-CDM scenarios. The curves show example scenarios with different $\mFDM$ and $\ffdm$ values, computed using the fitting formula
    of Equation \ref{eq:Rfitting} and the conditional luminosity function from Equation \ref{eq:phiLZ_disc}. The data points are the same as in Fig.~\ref{image_LF}.}
   \label{image_LFmodel}
\end{figure*} 

%%%%%%%%%%%%%
\subsection{Sensitivity to Spurious Halo Thresholds}

The results in the previous sub-sections adopt threshold choices of $S_{\rm T} = 0.25$ and $O_{\rm T} = 0.3$ for identifying and excising spurious halos. Here we vary these thresholds to assess the sensitivity of our results to these choices.  

We show how $\mathcal{R}$ varies with $S_{\rm T}$ and $O_{\rm T}$ for pure FDM models (with each of $\mFDM = 1, 10$ and $20$ at $z=8$) in Fig.~\ref{image_SO-thre-full}.  Since halos with $S_{\rm T}$ or $O_{\rm T}$ below the threshold values are considered spurious, lowering the threshold choices reduces the number of FDM halos that are removed, leading to higher FDM halo abundances and smaller $\mathcal{R}$ values. For example, for $\mFDM = 1$, the halo mass scale where $\mathcal{R} \sim 100$ shifts by a factor of a few -- from $M = 1.3 \times 10^9\MSUN$ to $M=3.7 \times 10^9\MSUN$ -- for the most lenient cut shown in the figure with $S_{\rm T}$ and $O_{\rm T}$ both $= 0.1$, compared to the most stringent case with $S_{\rm T}$ and $O_{\rm T} = 0.4$. However, the sensitivity to threshold choice is confined to high $\mathcal{R}$ portions, with the $\mFDM = 1$ results showing little dependence on $S_{\rm T}$ and $O_{\rm T}$ for $\mathcal{R} \lesssim 10$.

Furthermore, in the $\mFDM = 10$ and $\mFDM = 20$ models (middle and bottom panels), our simulation results reveal little sensitivity to $S_{\rm T}$ and $O_{\rm T}$. Note, however, the differing $y$-axis range of the three panels. The insensitivity to threshold choice in the higher $\mFDM$ scenarios is likely a consequence of the finite mass resolution of our simulations. That is, the simulations do not resolve the lower mass halo range where $\mathcal{R}$ is sufficiently large for the spurious halo thresholds to be important.

In summary, while the somewhat arbitrary spurious halo thresholds affect the HMF ratios,  this impact is largely confined to the strongly suppressed regime, $\mathcal{R} \gtrsim 10$. In practice, the detailed behavior of the suppression factor in this regime is of limited relevance for comparing with, e.g., UVLF data. Suppression at the level of $\mathcal{R} \gtrsim 10$ on well-probed halo mass scales is typically already strongly disfavored. Therefore, observational constraints are generally more sensitive to the shape and mass dependence of $\mathcal{R}$ at more moderate suppression levels. In this important regime we expect our results to be robust to the precise choice of spurious halo cuts adopted. 

Uncertainty bands for $dn/d{\rm ln}{M}$ are also shown in Fig. 4 of S2016 (for $f_{\rm F}=1$), reflecting $20\%$ variations around fiducial $S_\mathrm{T}$ and $O_\mathrm{T}$ values. These bands validate the robustness of those results to spurious halo threshold choices. Our wider range of $S_\mathrm{T}$ and $O_\mathrm{T}$ similarly reinforces our $\mathcal{R}$ results for the observationally relevant parameter space.

We summarise the results of this section using Fig.~\ref{image_mall}, where we show $\mathcal{R}$ as a function of halo mass across all pure FDM and mixed scenarios simulated in this work at a representative redshift of $z=8$. The solid lines show that our new HMF fitting formula successfully describes this full range.

%%%%%%%%%%%%%%%%%%%%%%%%%%%
\section{Discussion}
\label{sec:dis}

Here, we discuss the implications of the FDM HMF fitting formula derived in this work for observations of the UVLF during Cosmic Dawn. The suppression of the faint-end UVLF in FDM models has been discussed extensively in earlier work such as S2016 and \citet{2017PhRvD..95h3512C, 2019MNRAS.488.5551N, 2024ApJ...976...40W, 2025MNRAS.538.1830S}. We extend that work in two ways. First, we apply our revised FDM HMF fitting formula, which is better calibrated to the currently viable regime of $\mFDM \gtrsim 10$. Second, we explore mixed FDM-CDM cosmologies with a range of FDM fractions $\ffdm$. These improvements allow us to refine predictions for the high-$z$ UVLFs and quantify the differences between our new results and those of earlier HMF models. 

\begin{figure*}
 \centering
    \includegraphics[width=\textwidth]{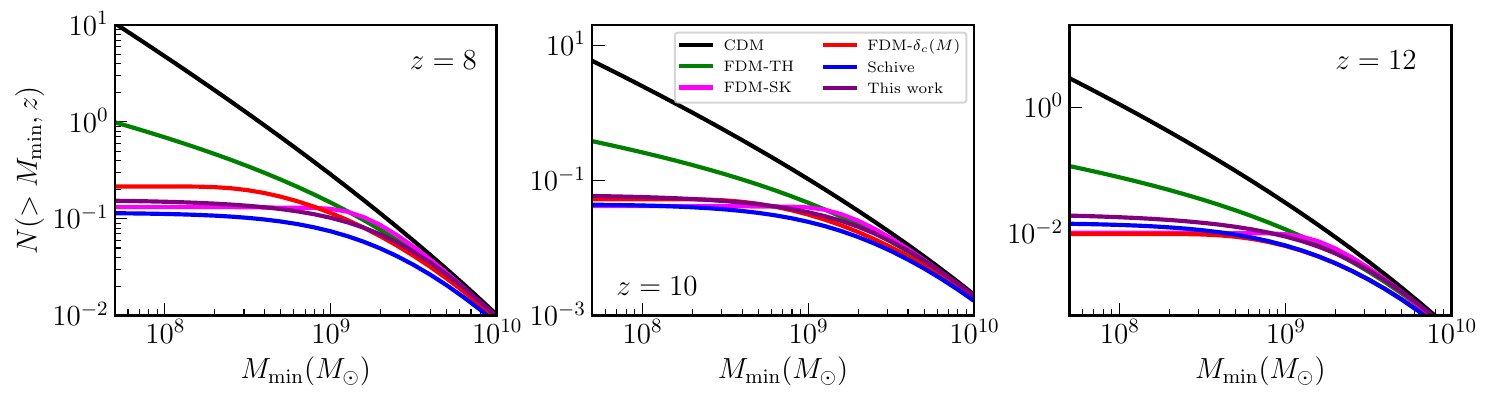}
    \caption{The cumulative abundance of dark matter halos for $\mFDM=5$ and $\ffdm=1$ at $z=$ 8, 10 and 12. The black curve is for CDM, while the other curves show different HMF model variants for FDM.}
   \label{image_HMF_cumu}
\end{figure*} 

\begin{figure*}
 \centering
\includegraphics[width=\textwidth]{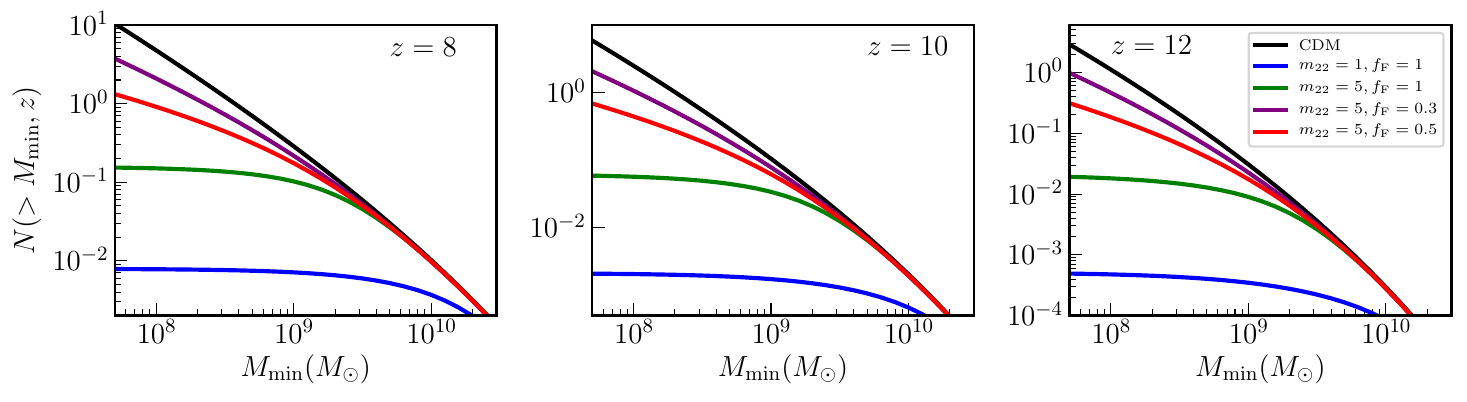}
    \caption{The cumulative abundance of dark matter halos at $z=$ 8, 10 and 12 for the different dark matter scenarios considered in this study. Here, we use the fitting formula of Equation \ref{eq:Rfitting}. }
   \label{image_HMFmodel_cumu}
\end{figure*} 

Specifically, we follow the UVLF modeling approach of \citet{2025MNRAS.538.1830S}, where the luminosity function is written as
\begin{equation}
    \phi(L,z) = \int_{M_{\rm min}}^\infty  \phi_c(L|M,z)\, \frac{d n(M, z)}{dM}\, \mathrm{d} M,
\label{eq:phiLZ_disc}
\end{equation}
with $\phi_c(L|M,z)$ specifying the conditional luminosity distribution for halos of mass $M$ at redshift $z$. We model $\phi_c$ as a lognormal distribution with scatter $\sigma_{\rm LN}=0.37$ and median luminosity
\begin{equation}
L_c(M,z)=L_0\left[\frac{(M/M_1)^p}{1+(M/M_1)^q}\right]\left(\frac{1+z}{7}\right)^r,
\end{equation}
where we adopt the best fit parameters from \citet{2025MNRAS.538.1830S}.\footnote{Following \citet{2025MNRAS.538.1830S}, we relate UV absolute magnitude to luminosity using $M_{\rm AB} = 51.6-2.5 \log_{10}\left[
L/({\rm~erg}~{\rm s}^{-1}~{\rm Hz}^{-1})\right].$}
Throughout this analysis, we keep the galaxy-halo connection fixed (i.e., we do not vary the conditional luminosity function model parameters) in order to isolate the impact of uncertainties in the FDM HMF itself. 

Fig.~\ref{image_LF} shows the UVLFs for different FDM HMF models with $\mFDM=5$ and $\ffdm=1$ over the redshift range $8\leq z \leq 12$. The interested reader can find further illustrative UVLF model results and measurements in Appendix \ref{sec:uvlf_add}. The figure shows that previous semi-analytic and simulation results span a wide range, especially at the faint end. For example, the ratios of CDM to FDM UVLFs for $z=8$ and $\mFDM=5$ at $\mathrm{M_{AB}=-12}$ (around the limiting faint-end magnitude measurable with JWST behind foreground cluster lenses, e.g. \citealt{Atek26}) range from $\approx 2.8-9.2$ across the scenarios considered. The minimum value of the CDM to FDM UVLF ratio is obtained for the FDM-TH HMF, while the FDM-Schive HMF yields the maximum ratio. This spread follows the trends seen for the HMFs in Section \ref{sec:hmf}. In this case ($z=8, \mFDM=5, \mathrm{M_{AB}}=-12$), our new fitting formula anchors the CDM to FDM UVLF ratio at 7.1, closest to the S2016 model but smaller than expected from that earlier work. Current UVLF measurements at $z\lesssim 10$ \citep[e.g.,][]{2021AJ....162...47B,2022ApJ...940...55B} are broadly consistent (within observational uncertainties) with both the S2016 fitting formula and our revised prescription, as shown in Fig.~\ref{image_LF}.

Since our results show a smaller suppression, it may slightly relax previous FDM parameter bounds placed assuming the S2016 formula \citep{2025MNRAS.538.1830S}. The semi-analytic model results are more discrepant with our new simulation-calibrated UVLF models, and are likely less reliable. The differences with earlier HMF results will become increasingly important as JWST improves the precision of the faint-end UVLF measurements and adds data at $z \gtrsim 10$.

We show the UVLFs for different combinations of $\mFDM$ and $\ffdm$ (using our HMF fitting formula) in Fig.~\ref{image_LFmodel}. Although we defer a full likelihood analysis to future work, the figure already gives a rough indication of the range in $\mFDM$ and $\ffdm$ parameter space which may be constrained. For example, the model with $\mFDM=5, \ffdm=1$ appears disfavored by the faint-end UVLF measurements at $z =8$, while the other cases in the figure are broadly compatible with the data. The $\mFDM = 0.1$ model case is roughly consistent with the observations, despite the low FDM particle mass, provided the FDM fraction is sufficiently low (e.g., $\ffdm \lesssim 0.1$). 

It would be interesting to compare mixed scenarios with upcoming JWST observations, especially if accurate and precise measurements around foreground lensing clusters can be achieved at $M_{\rm UV} \gtrsim -13$ and $z \gtrsim 10$. This would extend prior work \citep{2024ApJ...976...40W}, which employed Hubble Space Telescope high-$z$ UVLF measurements to test the mixed DM case, but using the real-space top-hat HMF with a modified $\sigma(M,z)$ (which gives a weaker suppression than our results).

We can also consider the cumulative abundance of DM halos, 
\begin{equation}
N(>M_{\mathrm{min}}, z) = \int_{M_\mathrm{min}}^\infty \frac{d n(M,z)}{dM}\, dM,
\end{equation}
which can be compared to the total abundance of galaxies, integrated over all observable luminosities. In \citet{2025MNRAS.538.1830S}, it was noted that the cumulative halo abundance saturates at non-zero $M_{\rm min}$ in FDM, and that this fact can be used to place a more model-independent bound on FDM particle mass, which is relatively insensitive to uncertainties in the galaxy-halo connection. Specifically, provided that each DM halo hosts one or fewer galaxies on average (see \citealt{2025MNRAS.538.1830S} for a discussion), then the cumulative halo abundance must match or exceed the integrated galaxy abundance. That is, models where the halo abundance falls short of the galaxy abundance can be ruled out. Of course, this technique is most constraining if measurements show high galaxy abundances. Therefore, robust measurements behind foreground lensing clusters, probing potentially abundant faint galaxies, are powerful in this regard. 

 The cumulative halo abundance for pure FDM models with $\mFDM=5$ at $z=8-12$ is shown in Fig.~\ref{image_HMF_cumu}. As with the differential UVLFs, there is a large spread in the model predictions with the ratio of the cumulative CDM to FDM HMFs at $M=10^9 M_\odot$ and $z=8$ varying between $\approx 1.95-3.9$. The minimum and maximum values are obtained for the FDM-TH HMF and FDM-Schive models, respectively. 

We compare the cumulative halo abundance for various pure and mixed FDM scenarios (assuming the HMF fitting formula of Equation \ref{eq:Rfitting}) in Fig.~\ref{image_HMFmodel_cumu}.
We note that the cumulative halo abundance is sensitive to $\mFDM$ and $\ffdm$, motivating a comparison of these models to upcoming measurements of the cumulative galaxy abundance during Cosmic Dawn.

%%%%%%%%%%%%%%%%%%%%%%%
\section{Conclusion}
\label{sec:con}

We presented a new suite of FDM and FDM-CDM simulations to investigate the HMF, and to explore the implications for forthcoming UVLF measurements during Cosmic Dawn. This work was, in part, motivated by the broad range of HMF model predictions in the current literature (see Fig.~\ref{image_hmfs}). For example, at a representative mass scale of $M = 10^9 M_\odot$ and $z=8$, the HMF models differ by more than a factor of two for $\mFDM=10$. 

We sought to better understand the spread in these predictions and to clarify which modeling approach is most accurate. For semi-analytic models, the HMF results are sensitive to the filtering scheme and barrier crossing treatment. The excursion set formalism may be less reliable in FDM cosmologies because the linear variance becomes nearly independent of halo mass below the suppression mass scale.  
On the other hand, a potential challenge for simulation measurements is that they require a careful removal of spurious halos sourced by discreteness noise in the initial conditions. Furthermore, previous simulation-based fitting formulas were calibrated for smaller $\mFDM$.

Our new simulations in pure FDM cosmologies broadly support the earlier simulation-based fitting formula of S2016, while refining it in several ways. First, the simulation-based HMFs agree substantially better with one another than with the semi-analytic models explored here. In particular, the simulated FDM HMF exhibits a stronger initial suppression than the sharp-$k$ space excursion set prediction, but a more gradual truncation at lower masses. Second, we find that uncertainties associated with spurious halo removal affect mainly the deeply suppressed regime where the FDM HMF is already reduced by an order of magnitude or more relative to CDM. This has a limited impact on current observational constraints, which are primarily sensitive to the more moderately suppressed regime. Finally, relative to S2016, our simulations favor a modestly weaker suppression, at the level of $\sim 15$--$30\%$ for $\mFDM \sim 10$--$20$ at representative halo masses and redshifts.

Our simulations also generalise previous work to handle mixed FDM-CDM scenarios, which are also theoretically well-motivated. Specifically, Equation \ref{eq:Rfitting} provides a new fitting formula, derived empirically from our simulations, which applies to both pure and mixed FDM-CDM cosmologies. Among other applications, this enables efficient UVLF predictions across a broad range of FDM scenarios, allowing direct comparisons with Cosmic Dawn era measurements from JWST, including recent measurements from \citet{Atek26}. 

Our UVLF analysis demonstrates that uncertainties in the FDM HMF propagate directly into predictions for the faint-end galaxy populations during Cosmic Dawn. For example, at $z=8$ with $\mFDM=5$ and $M_{\rm AB}=-12$, the ratio of CDM to FDM UVLFs spans a wide range, $\approx2.8\to 9.2$, between existing HMF prescriptions, while our revised fitting formula predicts an intermediate value of 7.1, 30\% weaker than the suppression inferred from the S2016 model. Similarly, the ratio of cumulative CDM to FDM halo abundances at $M=10^9\,M_\odot$ and $z=8$ varies between $\approx1.95$ and 3.9 depending on the HMF model adopted. Inspecting our new models, it appears that scenarios with $\mFDM \gtrsim 5-10$ can be stringently tested using faint-end UVLF measurements at Cosmic Dawn, while lower-mass mixed scenarios with small FDM fractions such as $\mFDM =0.1$, $\ffdm \sim 0.1$ can also be tested. A full MCMC analysis will be necessary to determine the quantitative bounds.  These efforts should soon be enhanced by new JWST measurements, with lensed galaxy observations at $\mathrm{M_{ AB}} \gtrsim-13$ and $z \gtrsim 10$ providing a powerful test of both pure and mixed FDM cosmologies.

In future work, it will be important to further test if our results are robust when the cut-off in the initial power spectrum of density fluctuations in FDM is augmented to include subsequent FDM dynamical effects. In principle, these could modify halo collapse and the HMF. Challenging wave-dynamical simulations across large cosmological volumes may be required to improve our understanding here. A more detailed investigation is also needed to determine whether the hint of redshift evolution (which is not captured by our fitting formula) in mixed scenarios (see Fig. \ref{image_m1f-new}) is physical or a numerical artefact, as this may have observational implications. Additional lower $\mFDM$ mixed scenario simulations will be helpful for understanding the transition between our HMF fitting formula and that of \citet{2025MNRAS.537..252D}.

The suppressed galaxy abundance at Cosmic Dawn will also modify the process of reionisation itself, which may be directly probed with other techniques. For example, the absence of small galaxies in FDM may require a larger role for bright galaxies or alternative ionizing sources in driving reionisation.  Future work combining JWST galaxy population measurements and 21-cm cosmology experiments such as HERA and the SKA could reveal astrophysical signatures of FDM or tighten existing bounds. 

%%%%%%%%%%%

\section*{Acknowledgements}
We thank Mustafa Amin, Frank van den Bosch, Xiaolong Du, Renee Hl\v{o}zek,  Alex Lagu\"{e}, D.J.E.~Marsh, Nathan Musoke, Ethan Nadler, Ravi Sheth, Tristan L. Smith, and Frank Wang for useful conversations. We acknowledge support from the Charles Kaufman Foundation through grant KA2022-129518. RG also acknowledges support from SERB, DST Ramanujan Fellowship no. RJF/2022/000141. We acknowledge the use of the {\texttt{Gadget} \citep{2005MNRAS.364.1105S}, \texttt{NgenIC} \citep{2005Natur.435..629S}, \texttt{NumPy} \citep{numpy}, \texttt{Matplotlib} \citep{matplotlib}} codes.

%%%%%%%%%%
\section*{Data availability}
The data underlying this article will be shared on request to the corresponding author. 

\newpage
\bibliographystyle{mnras}
\bibliography{notes}

%%%%%%%%%%%%%%%%%%%%%%%%%%%%%%%%%%%%%%%%%%%%%%%
\appendix

%%%%%%%%%%%
\section{Halo mass functions in a fuzzy dark matter model}
\label{sec:allhmf}

The halo mass function in the CDM model is \citep[e.g.,][]{1974ApJ...187..425P, 2002MNRAS.329...61S},
\begin{equation}
\left.\frac{dn}{d\ln M}\right|_{\mathrm{CDM}} = - f(\sigma) \frac{\rho_0}{M}  \frac{d\ln \sigma}{d\ln M}, 
\label{eq:ncdm}
\end{equation}
where $f(\sigma)$ is the fraction of the universe's mass contained in halos with $\nu \equiv\delta_{c}/\sigma(M,z)$, and $\sigma^2(M,z)$ is the variance of the linear density field at a mass scale $M$ and redshift $z$. Motivated by an ellipsoidal collapse model and tuned to fit $N$-body simulations, the Sheth-Tormen formula for $f$ is
\begin{equation}
f_\mathrm{ST}(\sigma) = A \sqrt{\frac{2 a}{\pi}} \left[1+ \left(\frac{\sigma^2}{a \delta_c^2}\right)^p \right] \frac{\delta_c}{\sigma} \exp \left(-\frac{a \delta_c^2}{2 \sigma^2}\right),
\label{eq:stf}
\end{equation}
with $A = 0.3222$, $a = 0.707$ and $p = 0.3$. The quantity $\rho_0$ is the present-day mean DM density. Here, $\delta_c = \delta_\mathrm{c,CDM}= 1.686$ is the mass and redshift-independent critical linear density contrast for spherical collapse. 

The variance of the linear overdensity field, on a smoothing scale $R$, can be written as
\begin{equation}
\sigma^2(R,z) = \frac{D^2(z)}{2\pi^2} \int_0^{\infty} k^2  P(k) W^2(kR) dk,
\label{eq:sigma}
\end{equation}
where $P(k)$ is the linear power spectrum of the matter density fluctuations at $z=0$. $D^2(z)$ is linear growth factor normalised to unity today, while $W(kR)$ is a smoothing function in Fourier space that filters out modes smaller than the scale $R$.

The linear power spectrum of the FDM model, $\PFDM(k)$, can be written as, $\PFDM(k) = \TFDM^2(k) \PCDM(k)$ \citep[see e.g.][]{2000PhRvL..85.1158H}, where $\TFDM$ is defined in the main text. There are a number of sensible choices for $W(kR)$, which are discussed below. In the case of CDM, detailed HMF predictions vary between these choices, but the general trends are robust.
In contrast, the optimal method for semi-analytic HMF predictions in FDM models is not yet known, as we now elaborate.

%%%%%%%%%%%%%%%%%%%%%%%%%%%%%%
\subsection{FDM HMF using real-space top-Hat filter with fixed barrier (FDM-TH-$\delta_c$)}

This prescription uses a spherical real-space Top-Hat filter, for which, 
\begin{equation}
W_\mathrm{TH}(k R) = \frac{3[\sin(kR) - kR \cos(kR)]}{(kR)^3}.
\end{equation}
The relation between the smoothing scale $R$ and the enclosed mass $M$ is $M=4\pi R^{3}\rho_{0}/3$. The method uses $\PFDM(k)$ in Equation (\ref{eq:sigma}), $\delta_c = 1.686$ in Equation~(\ref{eq:stf}), and yields an HMF of the same form as Equation~(\ref{eq:ncdm}), but with $\sigma(M,z)$ determined using $P_{\rm FDM}(k)$ with a real-space top-hat filter.

%%%%%%%%%%
\subsection{FDM HMF using Sharp-k Window Function (FDM-SK-${\delta_{c}}$)}
\label{sec:sharpkhmf}

In \citet{2021JCAP...01..051L} and \citet{2022MNRAS.510.1425K} the FDM HMF
was computed using a sharp-k window function, that is
\begin{equation}
W_\mathrm{SK}(R;k) = 
\left\{
\begin{split}
1 && \textrm{for } k \leq k_0\\
0 && \textrm{for } k > k_0,\\
\end{split}
\right.
\end{equation}
applying a constant critical density to analytically predict the FDM HMF. 
In contrast to filters defined in real space (such as the real-space top-hat), the sharp-$k$ filter is specified in Fourier space and does not correspond to a localised region in real space. Therefore, it does not naturally define a physical volume over which mass can be enclosed. Because of this, the smoothing scale $R$ cannot be uniquely translated into a halo mass $M$. In practice, an approximate $R$–$M$ relation is introduced using a calibration factor that is typically chosen to match results from numerical simulations. We adopt $k_0 = \alpha/R$ with $\alpha = 2.1$ and $\delta_c=\delta_\mathrm{c,CDM}$ to ensure that the FDM HMF matches with the CDM HMF at higher masses. We note that these choices are different from \cite{2022MNRAS.510.1425K}, where $k_0 = \alpha/R$ with $\alpha = 2.5$ and a rescaled  $\delta_c=1.396$ are used.

%%%%%%%%%%%%
\subsection{FDM HMF using real-space Top-Hat filter with mass-dependent barrier, FDM-TH-$\delta_c(M)$}
\label{sec:hmf-sa}

A semi-analytic prescription for the FDM HMF, as adopted in work such as \citet{2014MNRAS.437.2652M, 2015MNRAS.450..209B, 2016arXiv160505973M, 2017MNRAS.465..941D}, uses a mass-dependent barrier $\deltacrit(M)$ for halo formation, which causes a sharp cut-off in the HMF. A redshift-independent form of $\delta_c(M)$ used in previous work is \citep[see e.g.,][]{2016arXiv160505973M},
\begin{align}
\delta_c(M) ~ &\simeq ~ G_F(M) ~ \delta_\mathrm{c,CDM}, \nonumber \\ 
G_F(M) ~ &= ~ h_F(x)\exp(a_3 x^{-a_4}) \nonumber \\ &+ ~ [1 - h_F(x)]\exp(a_5 x^{-a_6}), \nonumber \\
x ~ &= ~ M/\MJeans^0, \\
h_F(x) ~ &= ~ \frac{1}{2} \left \{ 1 - \tanh\left[\frac{\MJeans^0}{h^{-1} \MSUN} (x - a_2)\right] \right \}, \nonumber
\\
\MJeans^0 ~ &= ~ a_1 \times 10^8 ~ \mFDM^{-3/2} \left ( \frac{ \Omegam h^2}{0.14} \right )^{1/4} h^{-1} \MSUN \nonumber,
\end{align}
with the fitting constants are
$\{ a_1, a_2, a_3, a_4, a_5, a_6 \} = \{ 3.4,~ 1.0,~ 1.8,~ 0.5,~ 1.7,~ 0.9 \}$.
The only difference between this HMF prescription from FDM-$\delta_c$ is the use of $\delta_c(M)$ instead of the mass-independent value of $\delta_c$ in Equation~(\ref{eq:stf}).

%%%%%%
\subsection{FDM HMF from simulations (FDM-Schive)}
\label{sec:hmf:schive}

In S2016, results are presented for several DM N-body simulations of FDM models using {\sc gadget2} \citep{2005MNRAS.364.1105S}, also providing a fitting formula for HMFs. Those simulations included the physics of FDM only through its initial matter power spectrum, followed by CDM-like growth. The fitting formula for the resulting modified HMF is \citep{2017PhRvD..95h3512C,2019MNRAS.482.3227N,2019MNRAS.488.5551N,2023MNRAS.524.4256M},
\begin{equation}
    \left.\frac{dn(M,z)}{d\ln M}\right|_{\rm SC,FDM} = \left.\frac{dn(M,z)}{d\ln M}\right|_{\rm CDM} \left[1+\left(\frac{M}{M_0}\right)^{-1.1}\right]^{-2.2},
    \label{eq:schivefitting}
\end{equation}
with $M_0=1.6\times10^{10}\mFDM^{-4/3}M_\odot$. In that work, the ratio of the FDM and CDM halo mass functions was found to be independent of z and dependent only on $\mFDM$. In this work, we explore the robustness of the above fit. We also simulate scenarios where FDM comprises only a fraction $f_{\rm F}\leq 1$ of the DM, and obtain fitting formulae appropriate to these scenarios.

%%%%%%%%%%%
\section{Additional UVLF Models and Data}
\label{sec:uvlf_add}

\begin{figure*}
 \centering
\includegraphics[width=\textwidth]{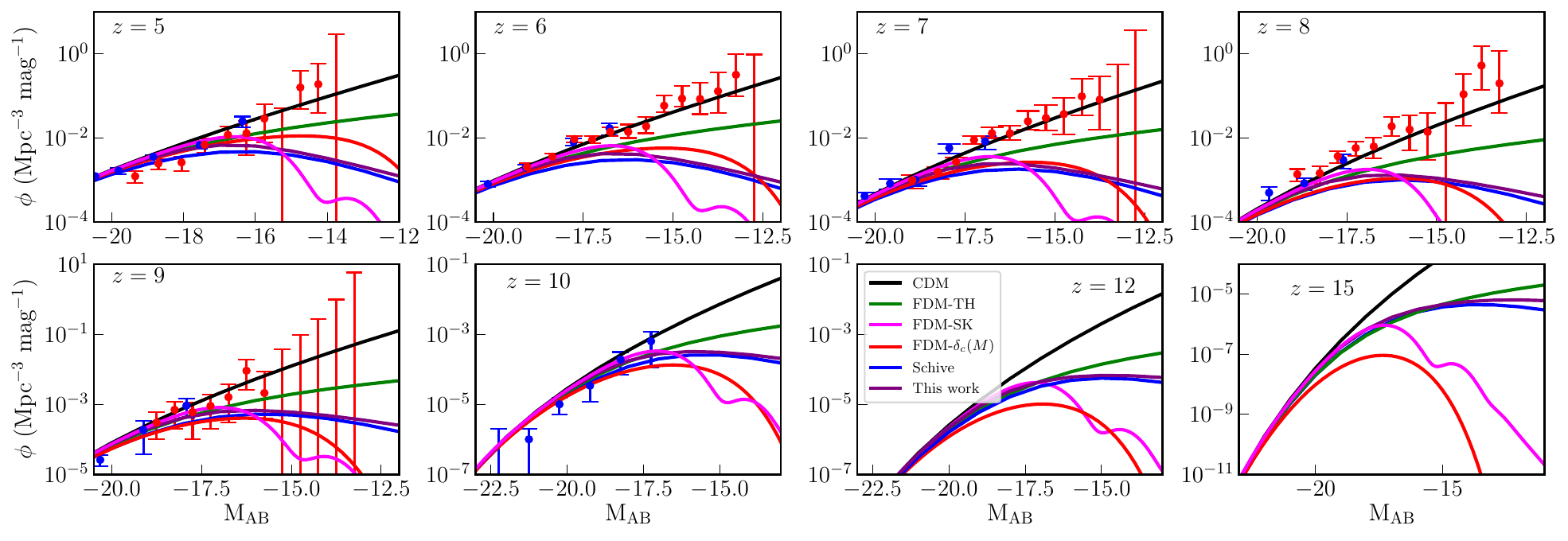}
    \caption{UVLFs for $\mFDM=1$ and $\ffdm=1$ at different redshifts between $z=5-15$. The black curve is for CDM, while the other curves show various FDM models. The points with $1\sigma$ error bars show the UVLF measurements between $z=5-10$ from \citet{2021AJ....162...47B} (blue, field galaxies) and \citet{2022ApJ...940...55B} (red, lensed galaxies).}
   \label{image_LF-app}
\end{figure*} 

\begin{figure*}
 \centering
\includegraphics[width=\textwidth]{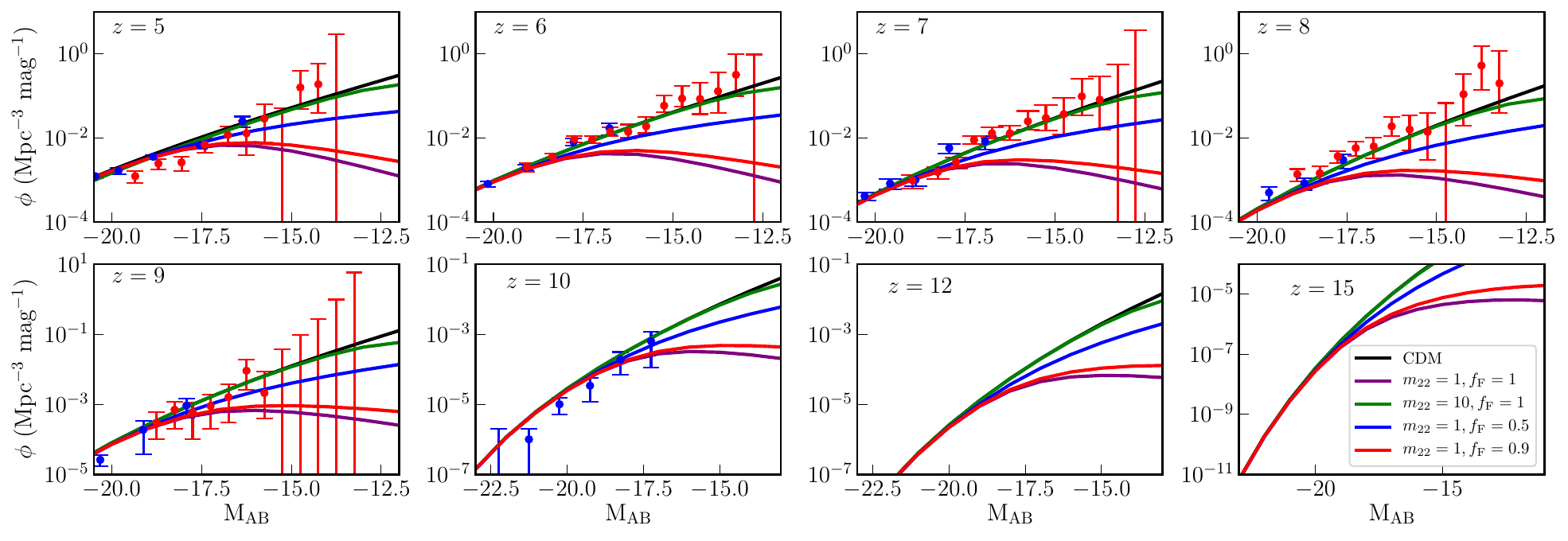}
    \caption{Similar to the previous plot, but here we show results for a range of different FDM model parameters.}
   \label{image_LFmodel-app}
\end{figure*}

\begin{figure*}
 \centering
    \includegraphics[width=\textwidth]{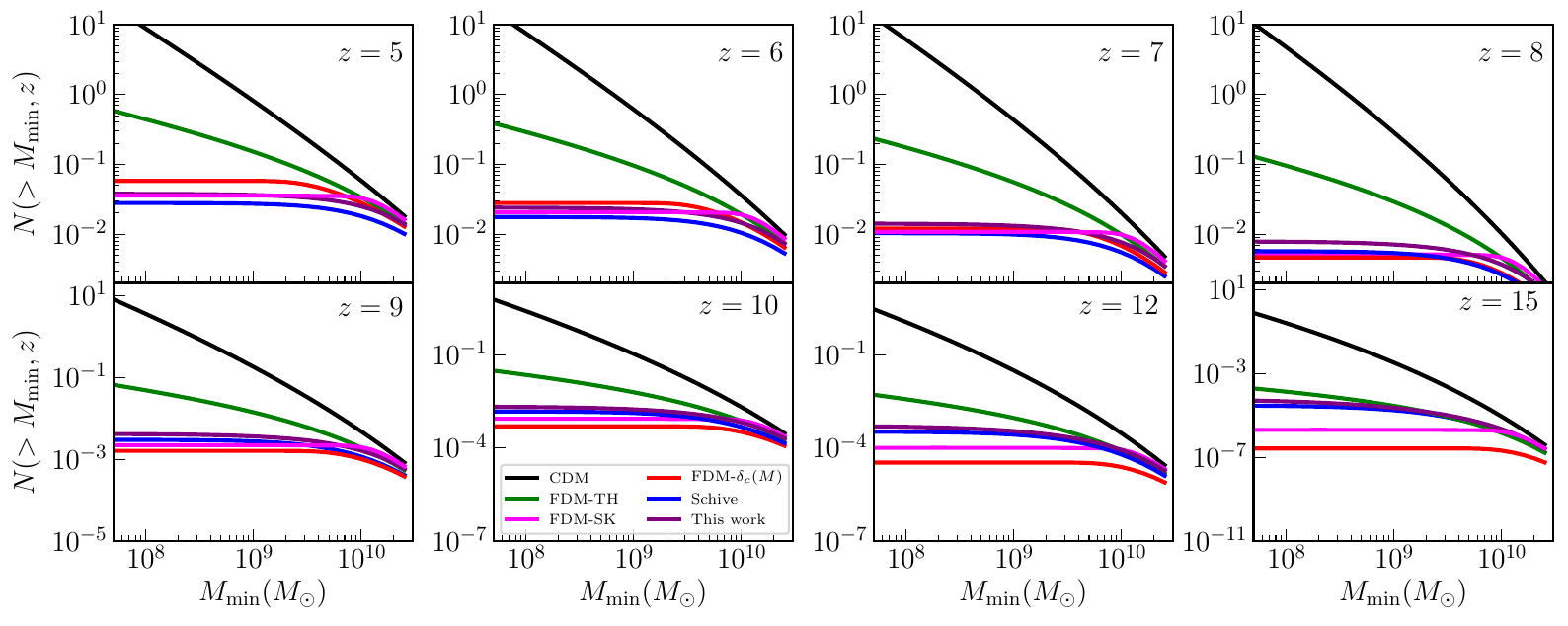}
    \caption{This plot is similar to Fig.~\ref{image_LF-app}, but here we show the cumulative halo abundance in models with $\mFDM=1$ and $\ffdm=1$.
    }
   \label{image_HMF_cumu-app}
\end{figure*} 

\begin{figure*}
 \centering
\includegraphics[width=\textwidth]{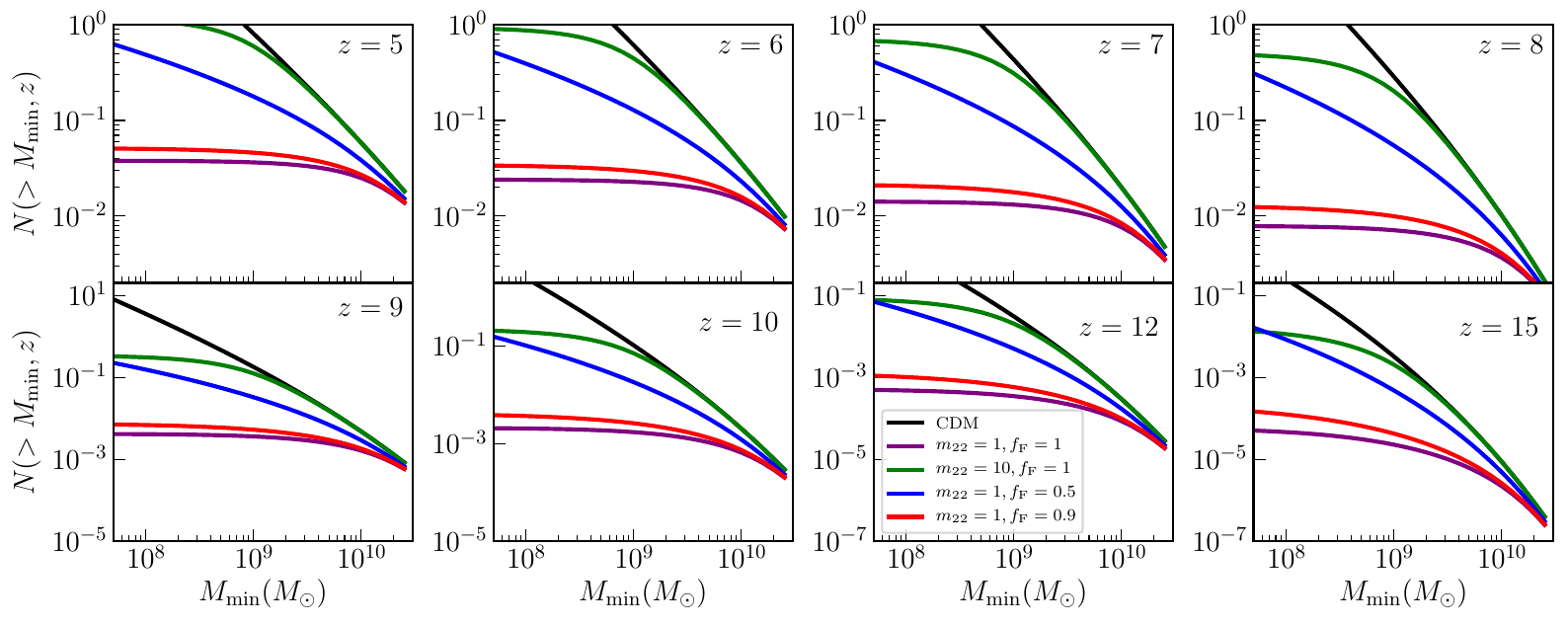}
    \caption{This plot is similar to Fig.~\ref{image_LFmodel-app}, but here we show the cumulative halo abundance for various FDM model parameters. 
    }
   \label{image_HMFmodel_cumu-app}
\end{figure*} 

For completeness, we include figures here for UVLF models and data, and for the cumulative halo abundance, over a broader range of redshifts and FDM parameters. See Figs.~\ref{image_LF-app}--\ref{image_HMFmodel_cumu-app}. These figures show similar trends to those in Section \ref{sec:dis}, but for an extended range in redshift and parameter space. 

%%%%%%%%%%%%%
\bsp	% typesetting comment
\label{lastpage}
\end{document}